\documentclass[a4paper,twocolumn,10pt]{article}

\usepackage{a4wide}
\usepackage{graphicx}
\usepackage{subfig}
\usepackage{amsmath}
\usepackage{algorithmic}
\usepackage{fixltx2e}
\usepackage{url}
\numberwithin{equation}{section}
\numberwithin{figure}{section}
\newcommand{\myWidth}[0]{\columnwidth}
\newcommand{\myHeight}[0]{0.2\textheight}

\begin{document}

\title{Investigating the Distribution of Password Choices}
\author{David Malone, Kevin Maher\\
Hamilton Institute,\\
 NUI Maynooth.\\}

\maketitle

\begin{abstract}
In this paper we will look at the distribution with which passwords
are chosen. Zipf's Law is commonly observed in lists of chosen
words.  Using password lists from four different on-line sources,
we will investigate if Zipf's law is a good candidate for describing
the frequency with which passwords are chosen. We look at a number
of standard statistics, used to measure the security of password
distributions, and see if modelling the data using Zipf's Law produces
good estimates of these statistics. We then look at the the similarity
of the password distributions from each of our sources, using guessing as a
metric. This shows that these distributions provide effective tools
for cracking passwords. Finally, we will show how to shape the distribution
of passwords in use, by occasionally asking users to choose a different
password.
\end{abstract}

\section{Introduction} 
\label{sec:intro}

In this paper we investigate if password frequency distribution can
be modelled by Zipf's Law. Zipf's Law is a probability distribution
where the frequency of an event is inversely proportional to its
rank on a frequency table. Here the rank of the most common event
is 1, the rank of the second most common is 2, and so on. Zipf's
Law has been observed when looking at the frequencies with which
words are used in natural language.  In our case, an event will
be the use of a particular password by a user.  To study this, we
use lists of users and passwords from hotmail.com, flirtlife.de,
computerbits.ie and rockyou.com. In each case, the list of usernames
and passwords were made public after a security incident. The lists
have 1800 users to 32 million users each.

There are a number of substantial differences between the choice
of passwords and the use of words in natural
language. In particular, passwords are usually chosen to be hard
to guess and there is a large literature of advice on how to choose
passwords (e.g. \cite{cert}). However, there are many reasons for
not choosing a password according to the recommendations: the advice
is arduous, it takes a long time to produce a compliant password,
the resulting password is likely to be a hash of letters and numbers
that will be hard to memorise. It has even been argued that in
general, the cost of choosing one of these passwords is far greater
than the cost of losing the information it is trying to protect
\cite{herley2009}.

Thus, rather than uniformly choosing from a list of non-dictionary
words the average user is likely to choose a password which they
can easily remember, such as their name, city where they live,
favourite team and so on which leads to certain passwords being
used more frequently than others. Since Zipf's Law has been observed
in many empirical data sets, we would like to investigate if it
provides a reasonable model for the passwords we see. To our knowledge,
this is the first paper to study if Zipf's Law applies to the choice
of passwords.

Seeing Zipf's Law, or any other distribution that is skewed in
favour of a number of passwords, has implications for security.  If
the right distribution of passwords can be identified, the cost of
guessing a password can be reduced. One naturally expects that,
say, the demographic of users of a site could be used to target an
attack; a website with a .ie domain is more likely to have Irish
themed passwords than a site with a .fr domain.

Zipf's law tells us that the number of occurrences of something is
inversely proportional to its rank on a frequency table, $y \sim
r^{-s}$, where $s$ is a parameter that is close to 1. By plotting
our datasets and fitting a value for $s$, we will see that, while a Zipf
distribution does not fully describe our data, it provides a
reasonable model, particularly of the long tail of password choices.
The heavy-tailed distribution of password choices could be used by
algorithm designers to more efficiently deal with passwords, such
those described in \cite{Schechter10}.

To establish if models, such as a Zipf distribution, can provide
useful predictions, we use metrics such as the guesswork \cite{pliam99}
and Shannon entropy. We calculate these metrics for both the fitted
model and the actual data, and compare the results. We find that
the actual metrics are within a factor of two of those predicted
by the Zipf distribution, and that the Zipf model usually provides
better predictions than a simple uniform model.

Another important question is how much one set of password choices
tells us about password choices in general. Consequently,
we compare the similarity of our different data sets using
guessing as a metric. We will show that by using common passwords
from one list, a speed up can be obtained when guessing the
passwords from another list.

Finally, we introduce a technique for making the passwords used by
users more uniform. When a user sets or resets a password, this
technique probabilistically asks them to choose a different password.
By basing this technique on the Metropolis-Hastings algorithm, we
can design it to produce a more uniform distribution of passwords
in use.

\section{Overview of Datasets}
\label{sec:interp}

We collected sets of passwords belonging to sites which were
previously hacked and the lists of passwords subsequently publicly
leaked. Since the sets
were gathered by different methods, such as key-logging, network
sniffing or database dumps, the lists may only contain a random,
and possibly biased,
sample of users. Our lists are from the 2009 hotmail.com list, the
2006 flirtlife.de list, the 2009 computerbits.ie list and the 2009
rockyou.com list.

Some of the lists also give multiple passwords for a small number
of users.  In this case, we cleaned up the sets by taking the
user's password as the last entry seen for that user, which would
hopefully correspond to a user initially typing the wrong password
and then typing the correct one, or in the case that the password
was changed, the most recent password. We also omitted any user
with a whitespace password. After the data was cleaned up, we
produced a table ranking passwords in order of decreasing frequency
of use by users.  Table~\ref{table:numseen} shows the number of
users and the number of distinct passwords for each set of data.
As is obvious from the table, for smaller lists there are relatively
more unique passwords.

\begin{table}%
\begin{center}%
{\small
\begin{tabular}{|l|r|r|r|}
\hline
Site & \#users & \#pass & $\displaystyle \frac{\mbox{\#pass}}{\mbox{\#users}}$\\
\hline
hotmail & 7300 & 6670 & 0.91 \\
flirtlife & 98930 & 43936 & 0.44 \\
computerbits & 1795 & 1656 & 0.92 \\
rockyou & 32603043 & 14344386 & 0.44 \\
\hline
\end{tabular}}	
\end{center}%
\caption{Number of users and number of distinct passwords seen for each site.}
\label{table:numseen}
\end{table}

Table~\ref{table:top10} summarises the top 10 passwords in each
list. We see that passwords such as \textbf{123456} and
\textbf{password} are very common.  The most common password,
``123456'' accounts for 0.7\% of the total passwords in the hotmail
data, 3.3\% in the flirtlife list and 2.0\% in the rockyou
list; ``password'' accounts for 1.2\% of the total passwords of
the computerbits list. This indicates that the password distribution
is skewed in favour of some common passwords.

The demographic of the users from each list is quite clear form the
first 10 passwords of the lists with each ccTLD. Note that the
hotmail.com data is believed to have been collected with phishing
targeted at the Latino community. The flirtlife list shows clear
signs of users speaking German and Turkish, and the computerbits
list contains place names of Irish interest. If we look at the data
from computerbits and rockyou we see that the name of the website
appears in the top ten of each list. It seems likely that this
method for choosing a password will also be used on other sites.

\begin{table*}%
\begin{center}%
{\small
\begin{tabular}{|l|l|r|l|r|l|r|l|r|}
\hline
Rank & hotmail& \#users & flirtlife& \#users & computerbits& \#users & rockyou& \#users\\ 
\hline
1 &123456    & 48 & 123456    & 1432 & password     & 20 & 123456    & 290729\\
2 &123456789 & 15 & ficken    &  407 & computerbits & 10 & 12345     &  79076\\
3 &111111    & 10 & 12345     &  365 & 123456       &  7 & 123456789 &  76789\\
4 &12345678  & 9  & hallo     &  348 & dublin       &  6 & password  &  59462\\
5 &tequiero  & 8  & 123456789 &  258 & letmein      &  5 & iloveyou  &  49952\\
6 &000000    & 7  & schatz    &  230 & qwerty       &  4 & princess  &  33291\\
7 &alejandro & 7  & 12345678  &  223 & ireland      &  4 & 1234567   &  21725\\
8 &sebastian & 6  & daniel    &  185 & 1234567      &  3 & rockyou   &  20901\\
9 &estrella  & 6  & 1234      &  175 & liverpool    &  3 & 12345678  &  20553\\
10&1234567   & 6  & askim     &  171 & munster      &  3 & abc123    &  16648\\
\hline
\end{tabular}}
\end{center}%
\caption{Top 10 Passwords for each list}
\label{table:top10}
\end{table*}

By taking obvious passwords and knowing the demographic of the users
at which the site is aimed one could build a comprehensive dictionary
that could be expected to cover the most common passwords in use
at a site. This implies that users, if they are concerned about
their accounts being hacked, should use less common passwords.  Even
something as simple as changing some of the characters to upper-case
or writing the word in `leet speak', could be expected to move the
password out of the most-common list. These results also confirms
something that system administrators have observed empirically,
that including a localised dictionary when checking password hashes
with crack will usually increase the number of recovered passwords.

\section{Distribution of Passwords}
\label{sec:dist}

{\small
\begin{figure}%
\centering
\subfloat[hotmail]{\includegraphics[width=\myWidth,height=\myHeight,keepaspectratio]{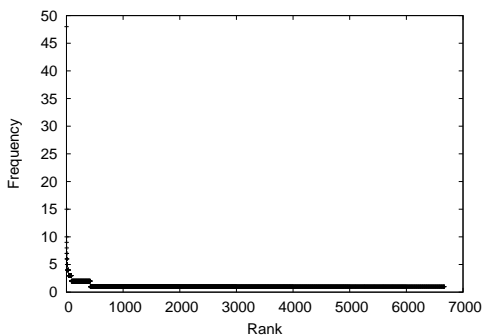}}\\%
\caption{Plot of rank vs frequency on a linear scale}%
\label{fig:linear}%
\end{figure}
}

{\small
\begin{figure}%
\begin{center}%
\subfloat[hotmail]{\includegraphics[width=\myWidth,height=\myHeight,keepaspectratio]{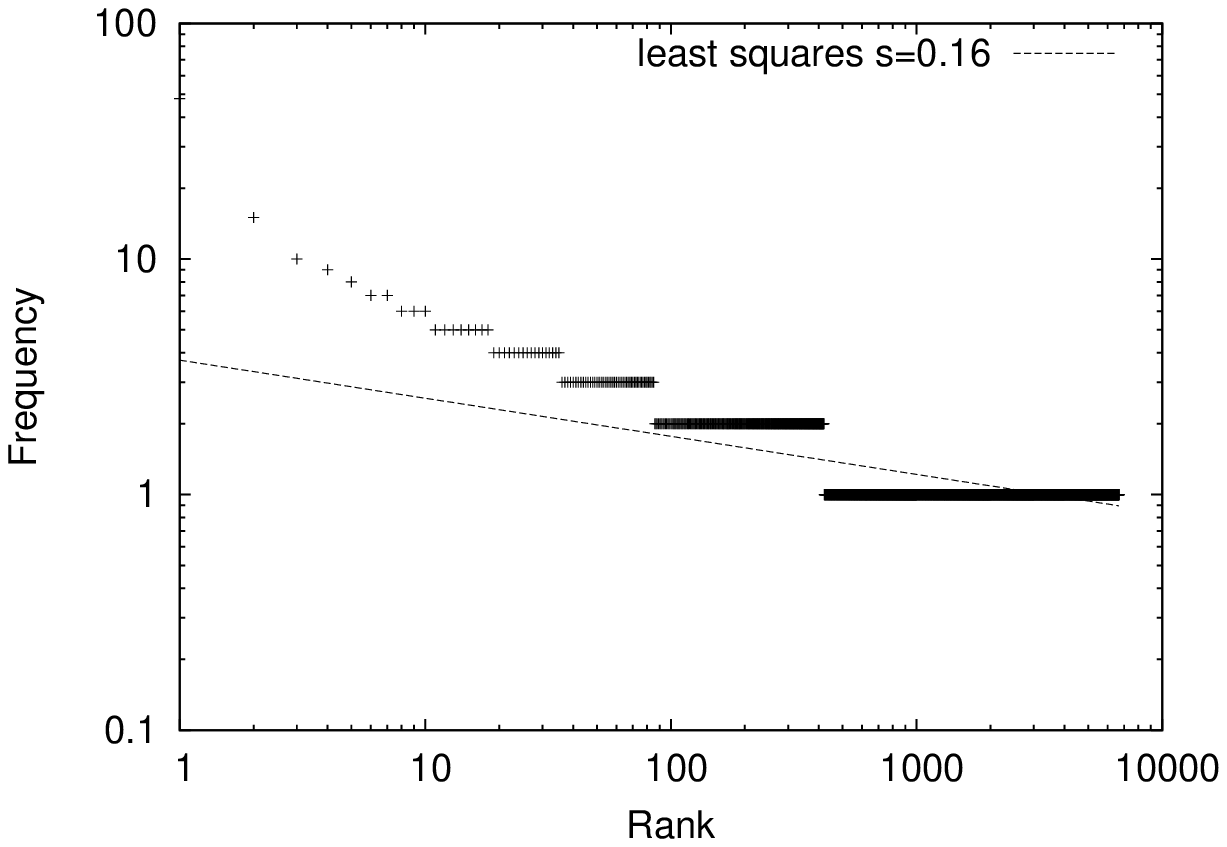}}\\\subfloat[flirtlife]{\includegraphics[width=\myWidth,height=\myHeight,keepaspectratio]{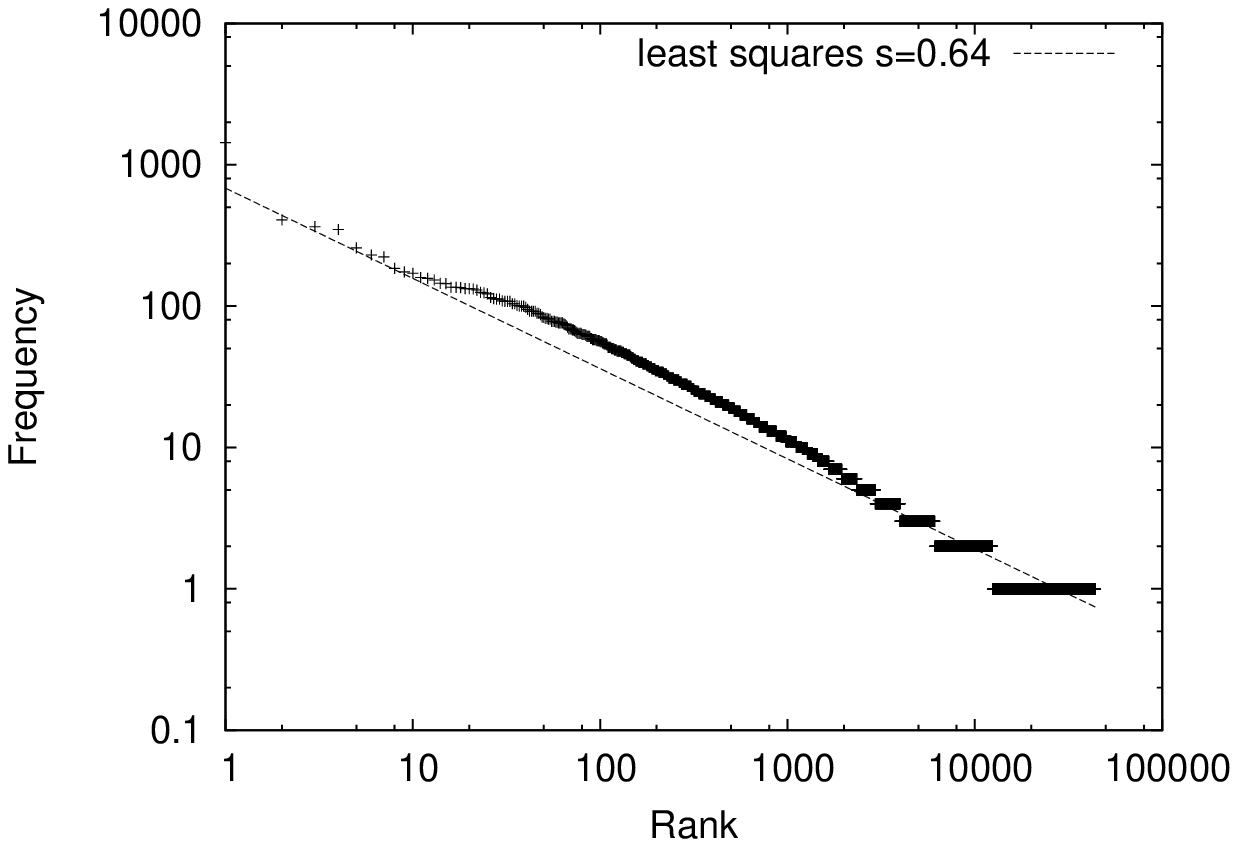}}\\\subfloat[computerbits]{\includegraphics[width=\myWidth,height=\myHeight,keepaspectratio]{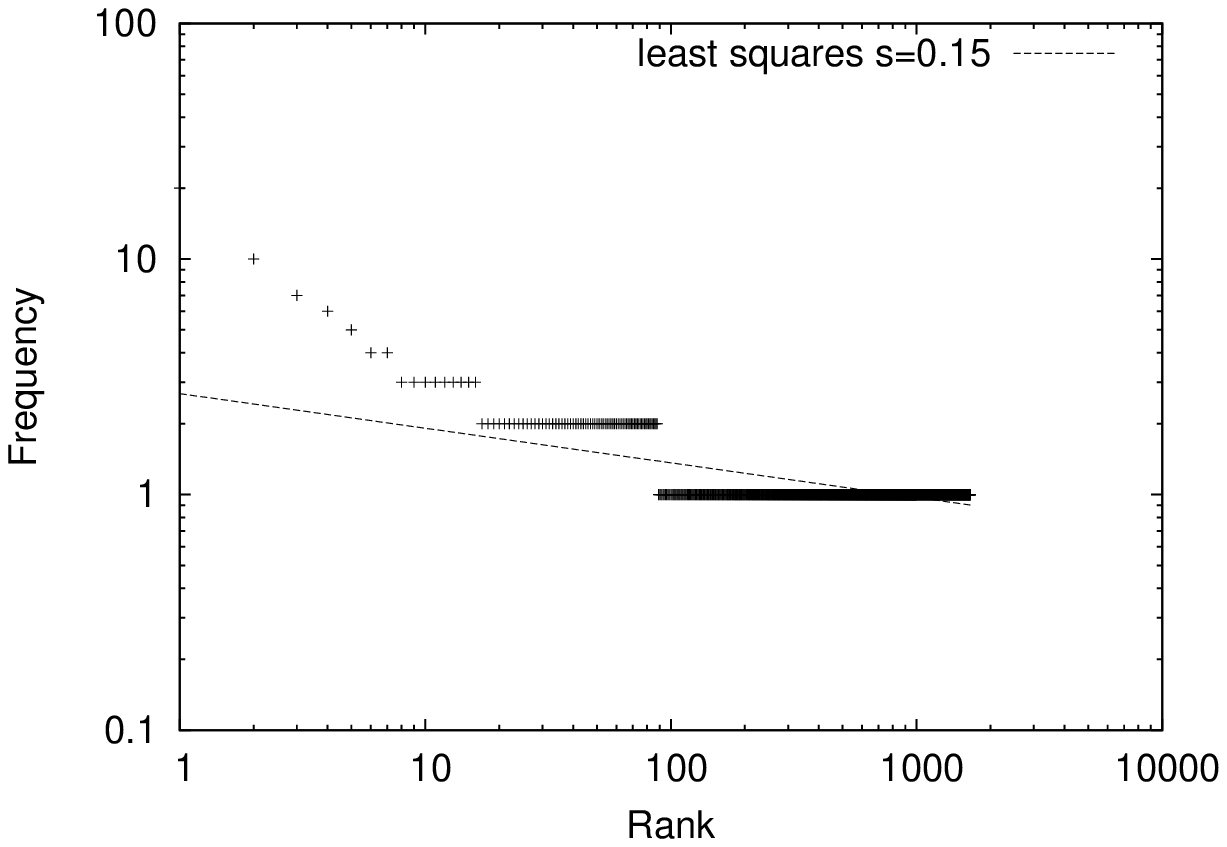}}\\\subfloat[rockyou]{\includegraphics[width=\myWidth,height=\myHeight,keepaspectratio]{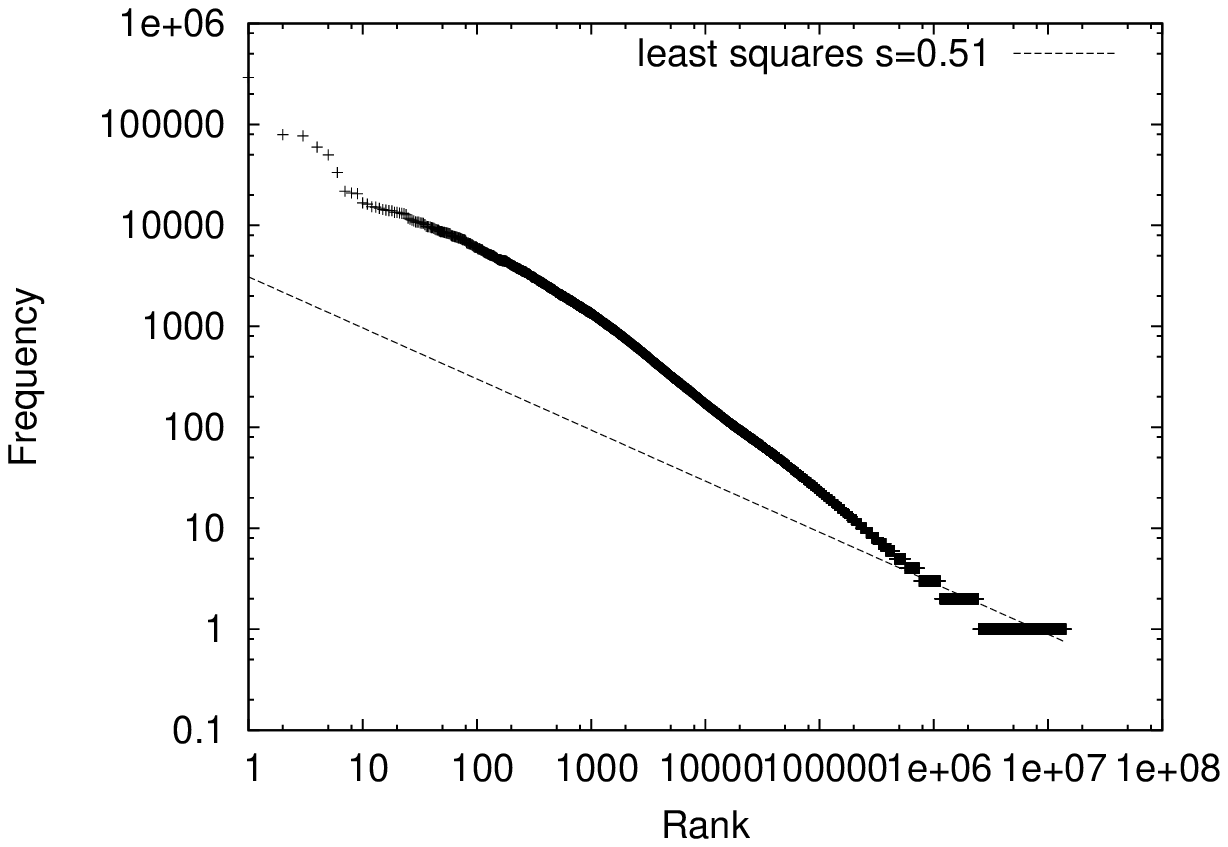}}\\\end{center}%
\caption{Plot of plot rank vs. frequency on log-log scale}%
\label{fig:logscale}%
\end{figure}
}

In this section we will look at how passwords are distributed in
our lists, and see how well these distributions match a Zipf model.
We will be interested in the frequency $f_i$ with which we see the
$i^{\mbox{\scriptsize th}}$ most popular password. Where passwords
are seen equal numbers of times, we break the tie randomly.

Fig.~\ref{fig:linear} shows the rank vs. frequency of our data
plotted on a linear scale for the hotmail list. There are a small
number of passwords with a high frequency, and many passwords with
a frequency of 1 or 2, which makes the graphs hard to read. Instead,
we plot the frequency versus rank on a log-log scale in
Figure~\ref{fig:logscale}. These graphs certainly show evidence of
heavy-tailed behaviour, with the frequency dropping much more slowly
than exponential. A Zipf distribution would appear as a straight
line on a log-log plot, where the parameter $s$ is the negative of
the slope.

If we fit a least-squares line to this data, as shown in
Figure~\ref{fig:logscale}, we get a slope which is too shallow because
a large fraction of the points have frequency 1 or 2, which biases
the slope towards 0. To account for this, we follow the method in
\cite{adamic-zipf} and bin the data logarithmically, as shown in
Figure~\ref{fig:logbin}. Here, we sum the frequency of all ranks
between $2^n$ and $2^{n+1}-1$.  We see that this gives us a slope
which better fits our data, and that the line appears a relatively
good fit. We use this binned slope as a basis for modelling our
data with a Zipf distribution.

{\small
\begin{figure}%
\begin{center}%
\subfloat[hotmail]{\includegraphics[width=\myWidth,height=\myHeight,keepaspectratio]{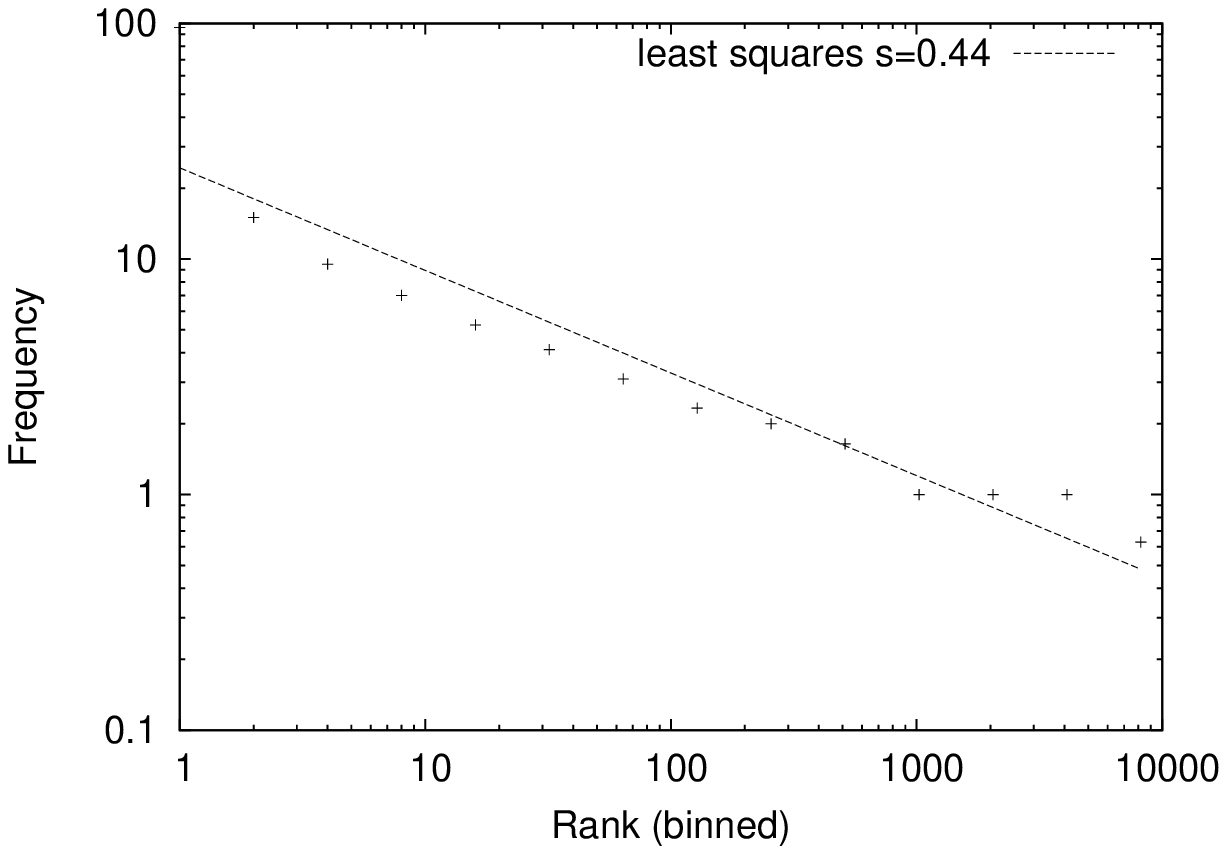}}\\\subfloat[flirtlife]{\includegraphics[width=\myWidth,height=\myHeight,keepaspectratio]{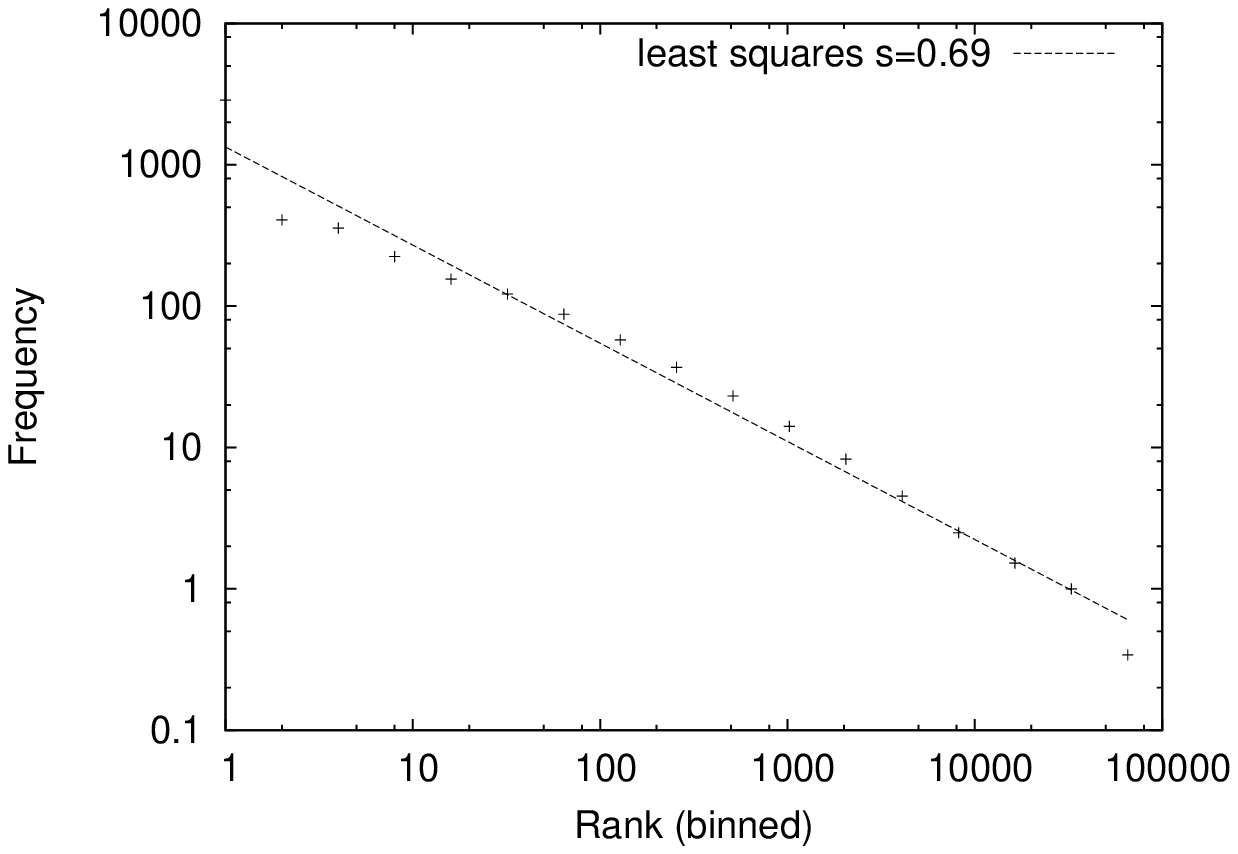}}\\\subfloat[computerbits]{\includegraphics[width=\myWidth,height=\myHeight,keepaspectratio]{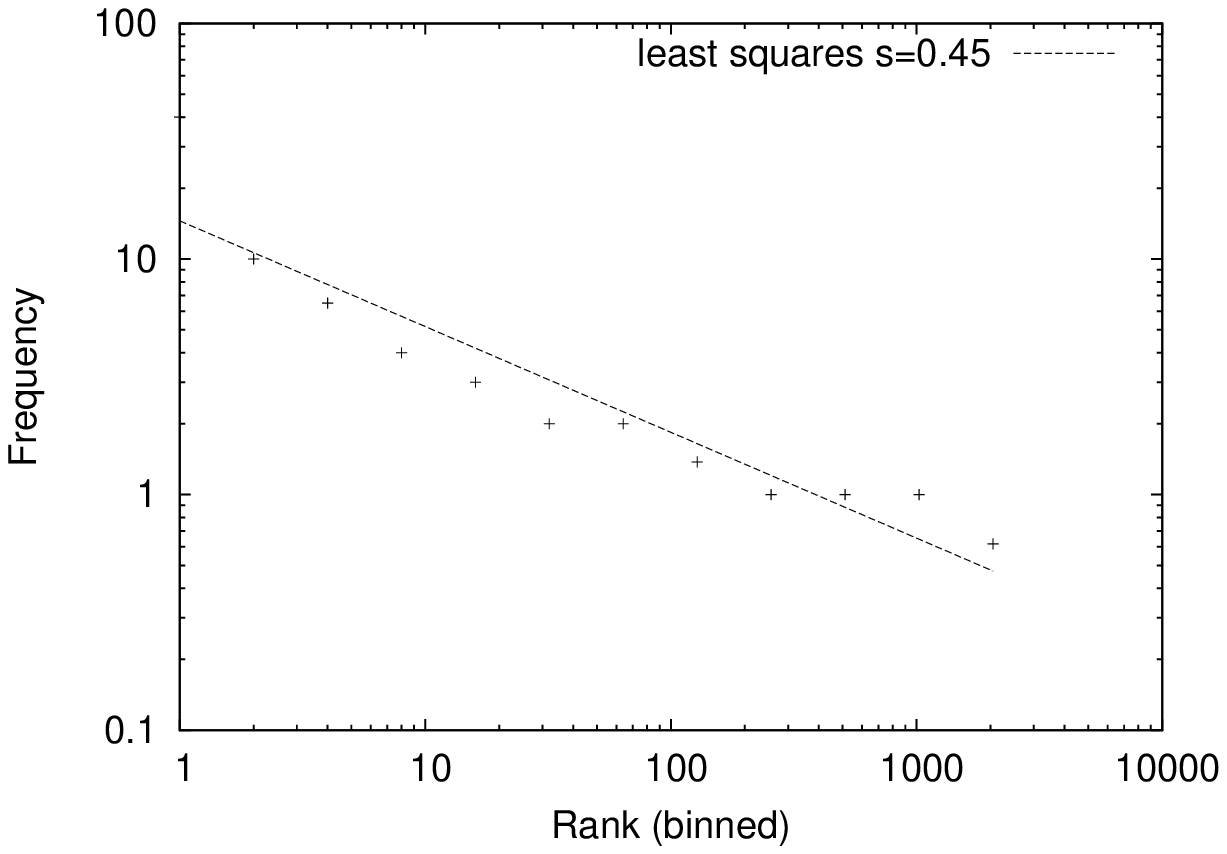}}\\\subfloat[rockyou]{\includegraphics[width=\myWidth,height=\myHeight,keepaspectratio]{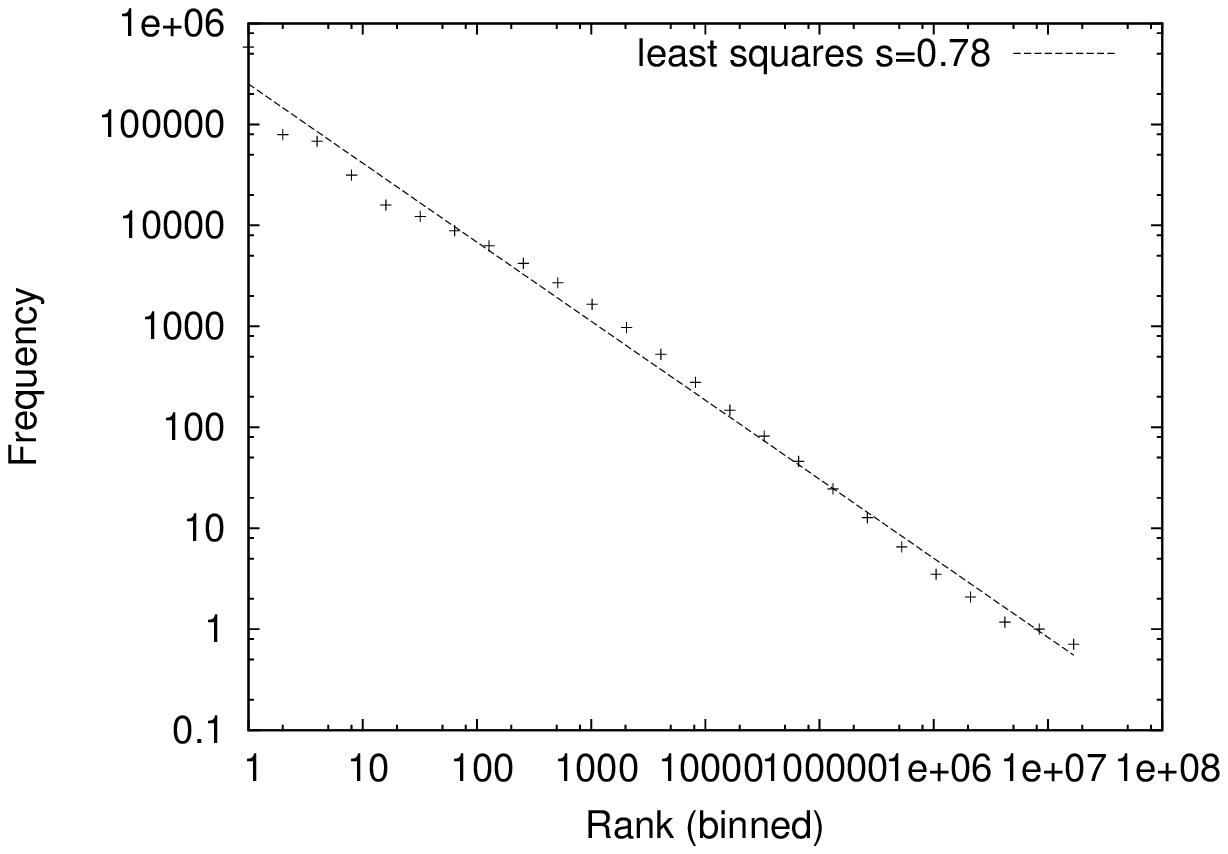}}\\\end{center}%
\caption{Exponentially binned plot of Figure~\ref{fig:logscale}}%
\label{fig:logbin}%
\end{figure}
}

An alternate way to view the data is to look at the number of
passwords $n_k$ that are each used by exactly $k$ users. We plot
this in Figure~\ref{fig:nkfit} on a log-log scale. As explained in
\cite{adamic-zipf}, if the data is Zipf-distributed, we expect this
graph to also be a straight line with slope $-(1+1/s)$.
As we can see, we also need to bin this data before fitting a line,
and the result it shown in Figure~\ref{fig:nkbin}. Again, we see a
line is a relatively good fit, with the largest discrepancy appearing
for computerbits, the smallest list. The resulting slopes are
summarised in Table~\ref{tab:slopes}.

{\small
\begin{figure}%
\begin{center}%
\subfloat[hotmail]{\includegraphics[width=\myWidth,height=\myHeight,keepaspectratio]{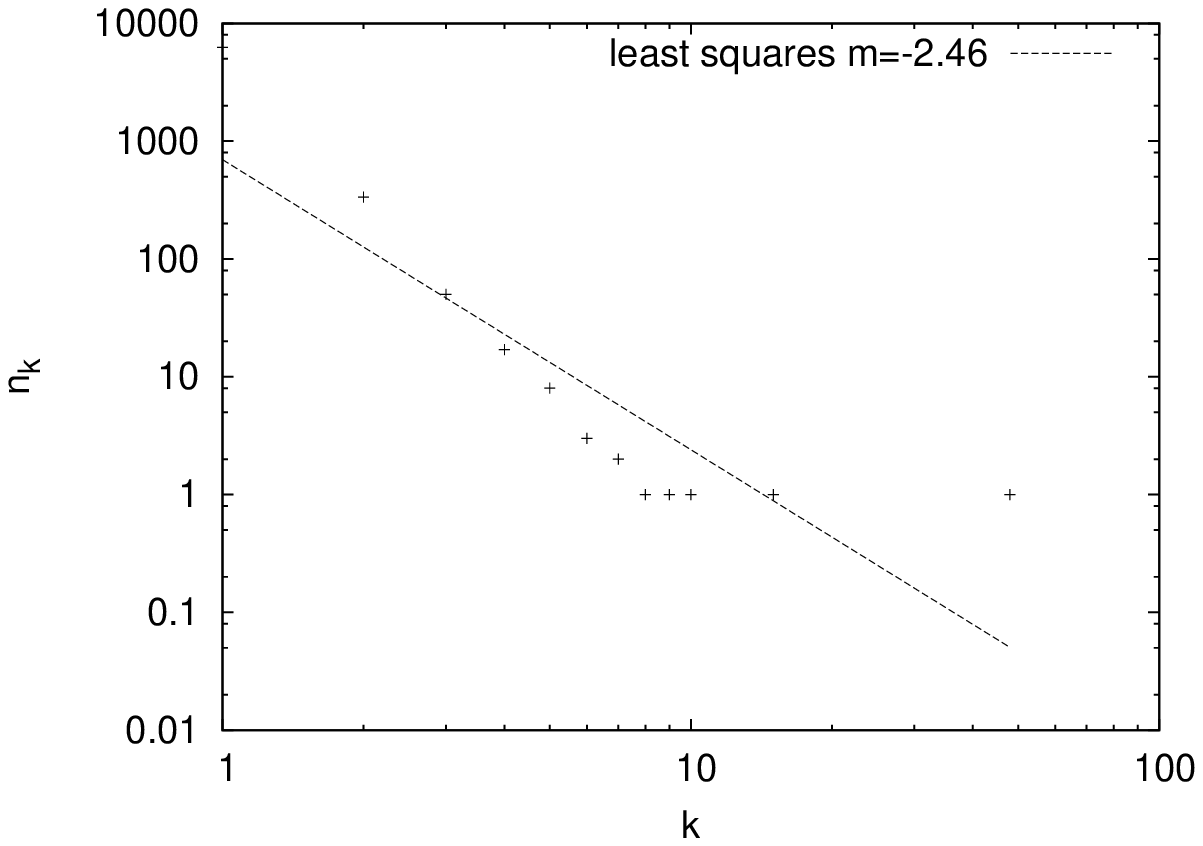}}\\\subfloat[flirtlife]{\includegraphics[width=\myWidth,height=\myHeight,keepaspectratio]{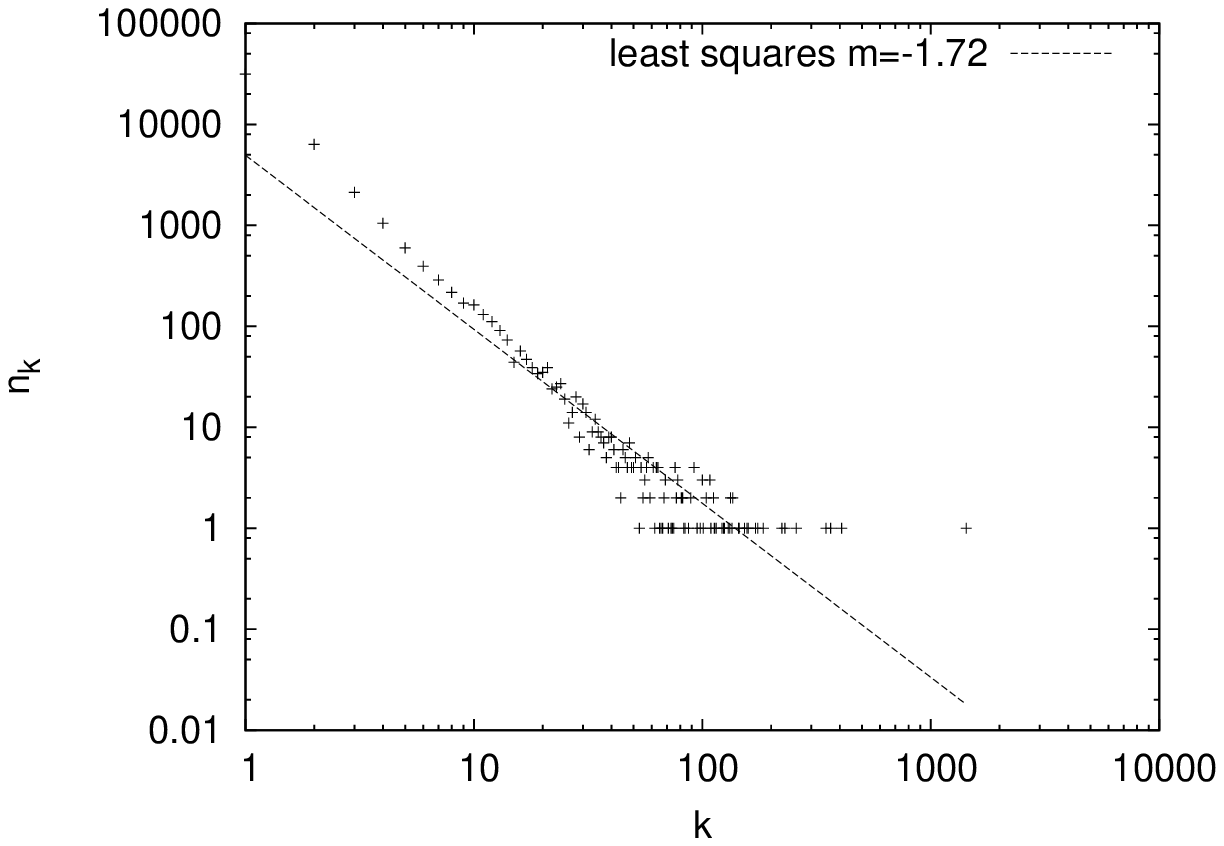}}\\\subfloat[computerbits]{\includegraphics[width=\myWidth,height=\myHeight,keepaspectratio]{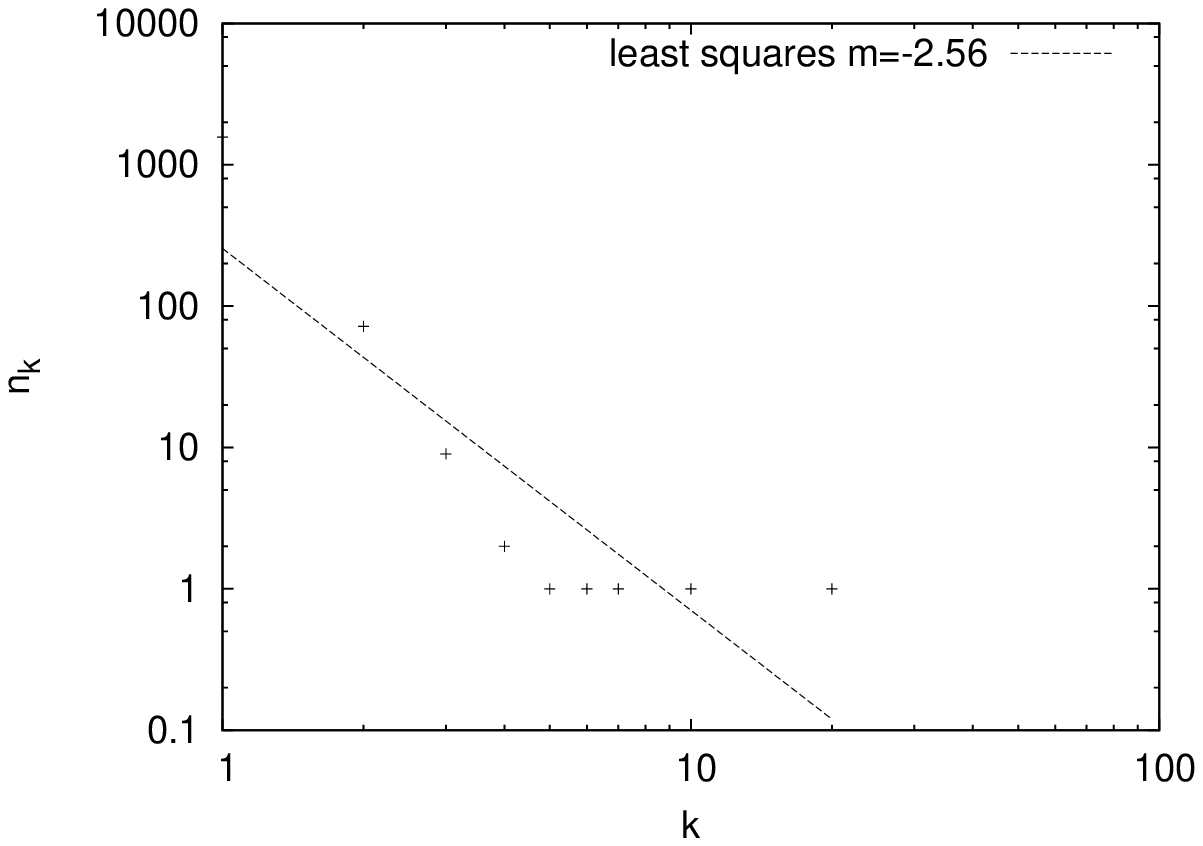}}\\\subfloat[rockyou]{\includegraphics[width=\myWidth,height=\myHeight,keepaspectratio]{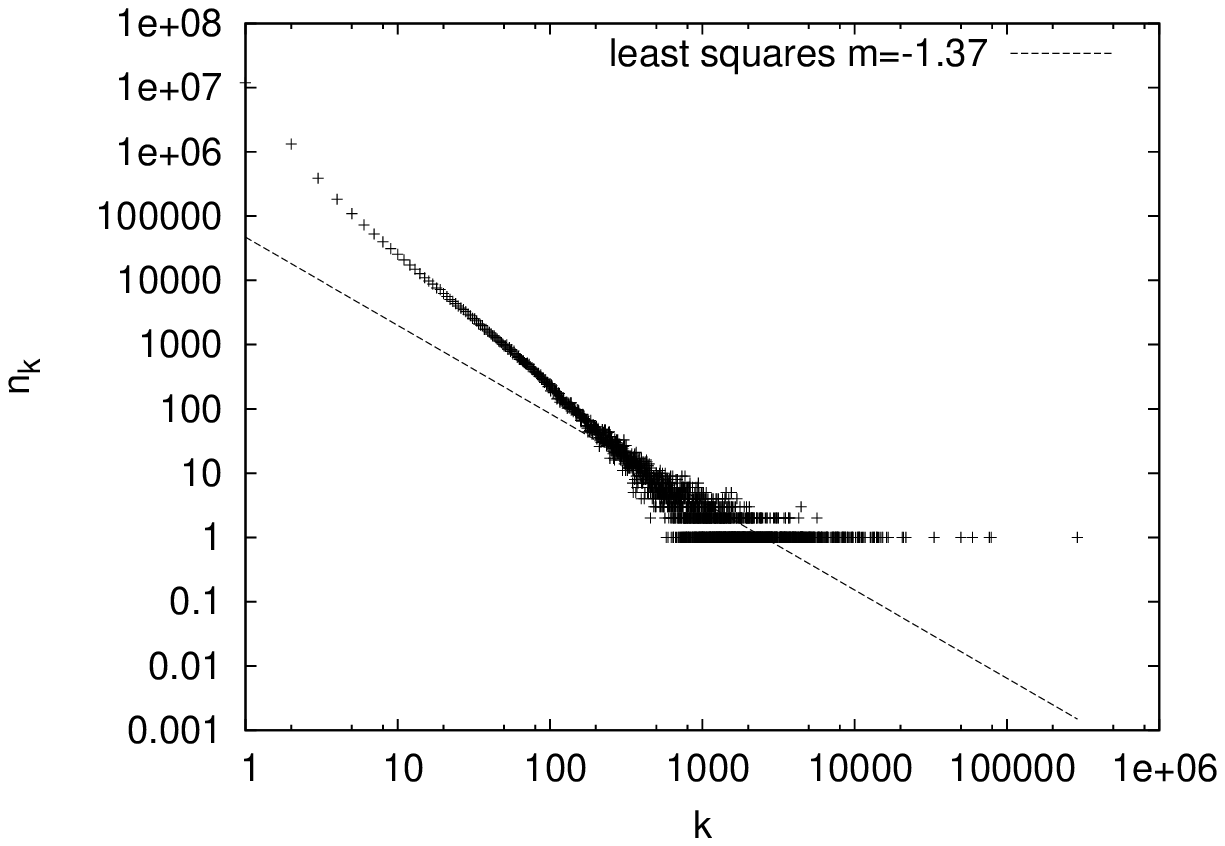}}\\\end{center}%
\caption{Plot of $k$ vs. $n_k$ on log-log scale.}%
\label{fig:nkfit}%
\end{figure}
}

{\small
\begin{figure}%
\begin{center}%
\subfloat[hotmail]{\includegraphics[width=\myWidth,height=\myHeight,keepaspectratio]{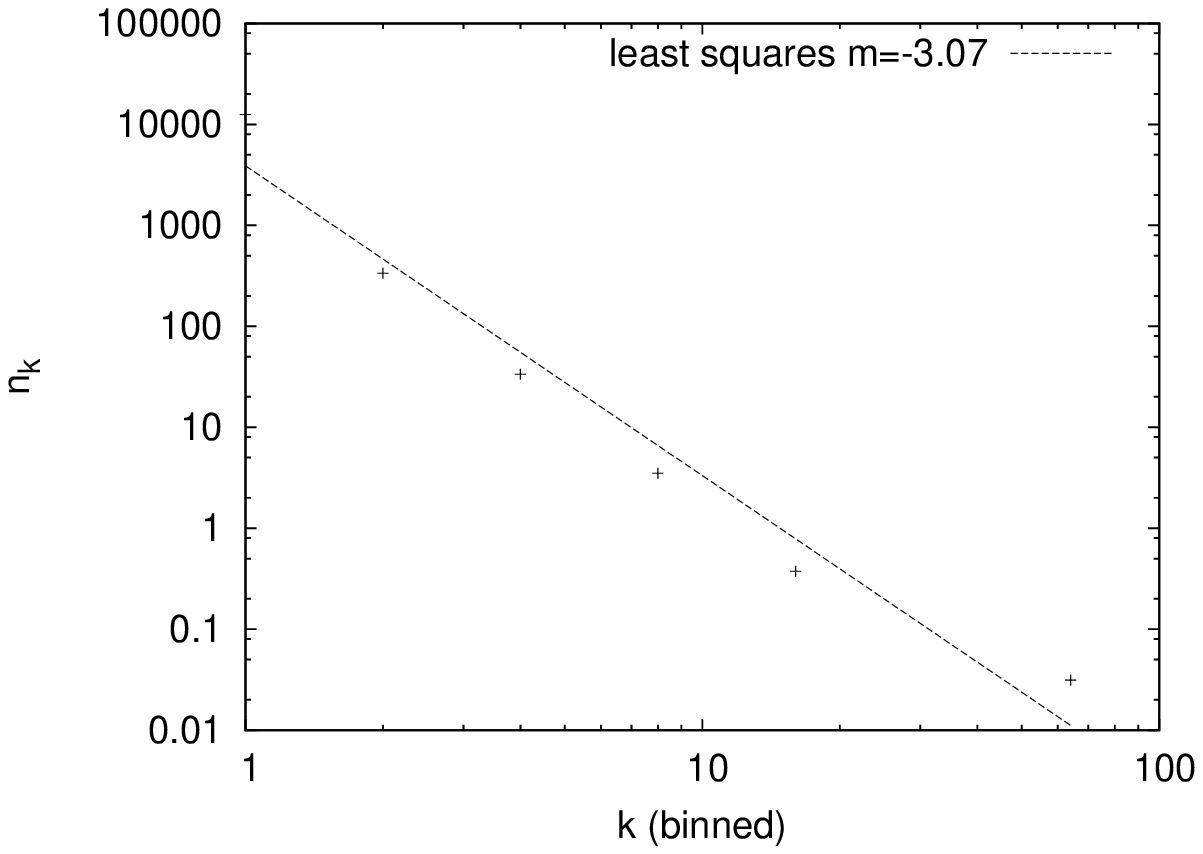}}\\\subfloat[flirtlife]{\includegraphics[width=\myWidth,height=\myHeight,keepaspectratio]{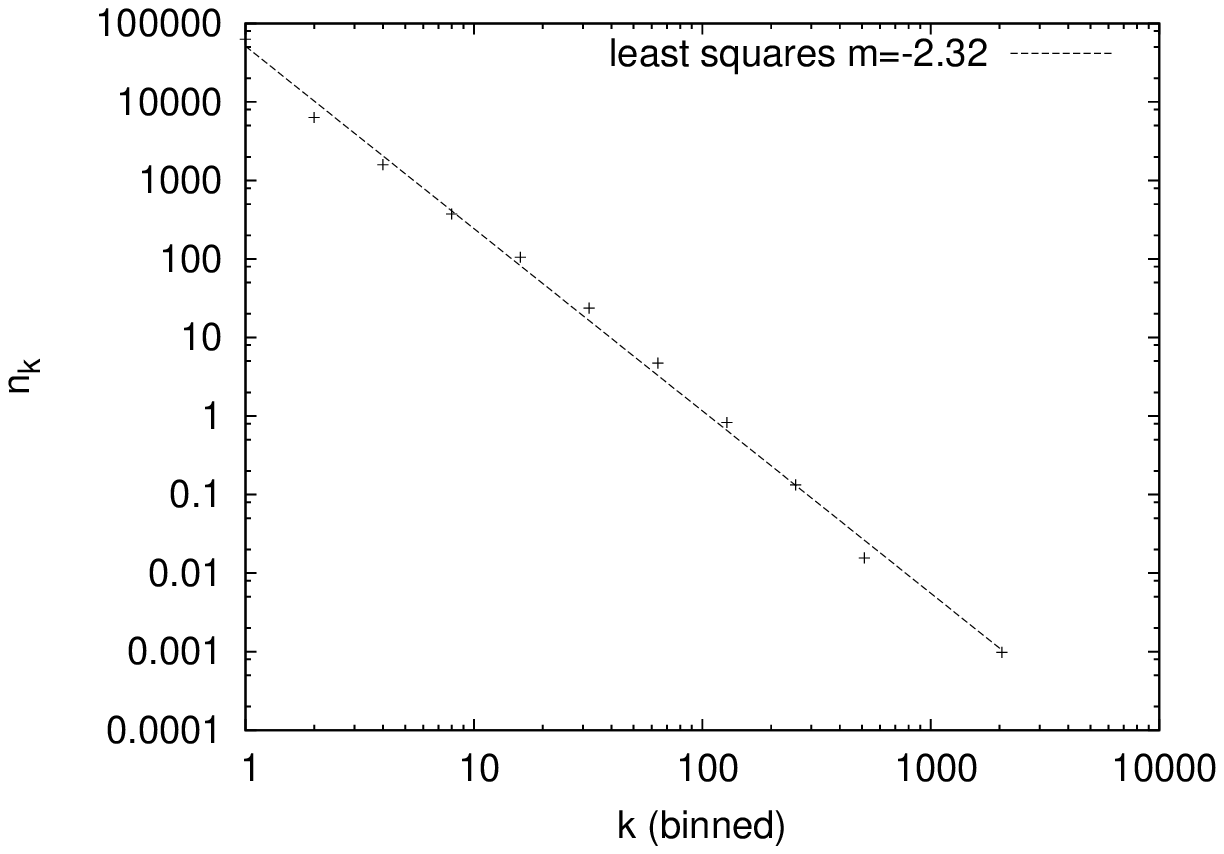}}\\\subfloat[computerbits]{\includegraphics[width=\myWidth,height=\myHeight,keepaspectratio]{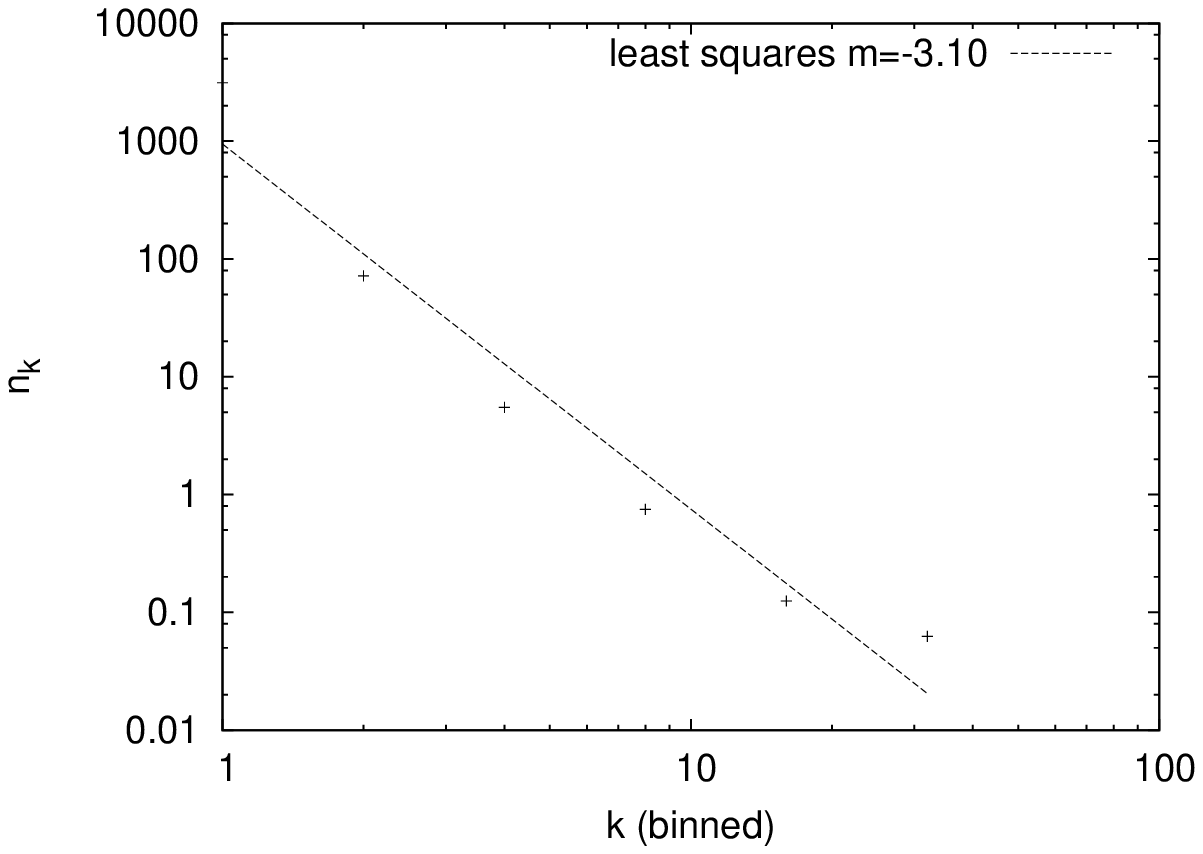}}\\\subfloat[rockyou]{\includegraphics[width=\myWidth,height=\myHeight,keepaspectratio]{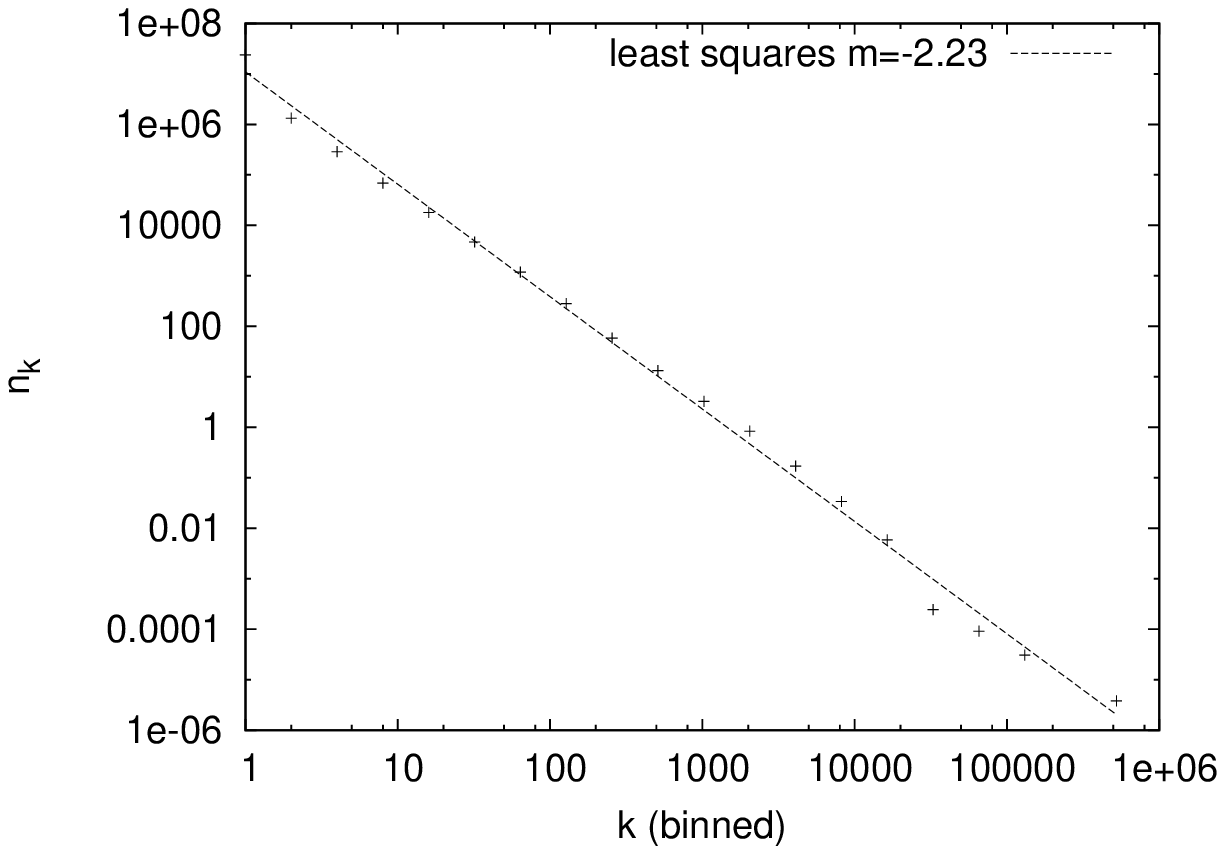}}\\\end{center}%
\caption{Plot of $k$ vs. $n_k$, exponentially binned on log-log scale.}%
\label{fig:nkbin}%
\end{figure}
}

\begin{table}%
\begin{center}%
{\small
\begin{tabular}{|l|r|r|r|r|}
\hline
         & hotmail & flirtlife & c-bits & rockyou \\
\hline
$s$      & 0.16    & 0.64      & 0.15         & 0.51    \\
raw      &         &           &              &         \\
\hline
$s$      & 0.44    & 0.69      & 0.45         & 0.78    \\
binned   &         &           &              &         \\
\hline
$-m$     & 2.46    & 1.72      & 2.56         & 1.37    \\
raw      &         &           &              &         \\
\hline
$-m$     & 3.07    & 2.32      & 3.10         & 2.23    \\
binned   &         &           &              &         \\
\hline
$1+1/s$  & 3.27    & 2.45      & 3.22         & 2.28    \\
binned   &         &           &              &         \\
\hline
\end{tabular}
}
\end{center}%
\caption{Parameters for Zipf distribution, estimated by least squares fits
to frequency and $n_k$ vs. $k$ graphs.}
\label{tab:slopes}
\end{table}

We can also build a maximum liklihood estimator (MLE) for a truncated
Zipf distribution, which assigns probability proportional to $r^{-s}$
to passwords with rank $r = 1\ldots N$. The MLE for $N$ is just the
number of passwords with non-zero frequency and the MLE for $s$ can
be constructed using standard techniques as described in \cite{clauset2009}.
This has the advantage of providing both estimates of the standard
error in $s$ and a p-value\footnote{P-values are calculated by
generating samples using a Zipf with the estimated parameters,
applying the same process of sorting and estimation. We then calculate
the fraction which exceed the Anderson-Darling modification of the
K-S statistic of our actual data.}. The results are shown in
Table~\ref{tab:mles}.

\begin{table}%
\begin{center}%
{\small
\begin{tabular}{|l|r|r|r|r|}
\hline
         & hotmail & flirtlife & c-bits & rockyou \\
\hline
$s$      & 0.246   & 0.695     & 0.23         & 0.7878  \\
MLE      &         &           &              &         \\
\hline
$s$      & 0.009   & 0.001     & 0.02         & $<0.0001$ \\
stderr   &         &           &              &         \\
\hline
p-value  & $<0.01$ & 0.57      & $<0.01$      & $<0.01$ \\
\hline
\end{tabular}
}
\end{center}%
\caption{Parameters for Zipf distribution, estimated by maximum liklihood.}
\label{tab:mles}
\end{table}

We see that the estimates for $s$ provided by the MLE for the
flirtlife and rockyou data are quite close to those provided by
least-squares estimate. The MLE estimates for $s$ for the smaller
data sets are between the binned values (around 0.45) and the raw
values (around 0.15). We see that the p-values indicate that the
hotmail, computerbits and rockyou data are unlikely to actually be
Zipf distributed. However, for the hotmail and computerbits data
the largest discrepancy between the Zipf's Law and the data is for
the first few passwords, indicating that the tail of the data could
pass for Zipf with a higher p-value.

To summarise, we have seen that the password frequency data has
heavy-tailed characteristics by plotting it on a log-log plot.  Both
least-squares and maximum-liklihood estimates indicate that if Zipf
distributed, the $s$ parameter is small. However, p-values indicate
the data is unlikely to be drawn exactly from Zipf's Law.

\section{Password Statistics}

In this section we will look at a number of statistics relevant to
passwords that can be derived from the distribution of how passwords
are chosen. We will look at these statistics when calculated directly
from our lists, and when calculated using two simple models of
the lists. For the real lists, we calculate our statistics
assuming the probability of the password of rank $i$ appearing is
$f_i/N$, where $f_i$ is the frequency with which we observed that
password and $N$ is the total number of passwords observed.

The first model assumes password choices are uniform over all
passwords seen, i.e., if the number of passwords is $N$ then a
password is chosen with probability $1/N$. The second model assumes
that password choices are distributed with a Zipf distribution,
i.e., the probability of password with rank $i$ being used is $P_i
= K i^{-s}$, where $s$ is the parameter found in Section~\ref{sec:dist}
and K is a normalising constant.

Now let us consider the statistics of interest. Some of these
statistics place considerable emphasis on the tail of the distribution,
and can be sensitive to relatively small gaps between the model and
the data. The first statistic is the \emph{guesswork}, which is the
mean number of guesses needed to correctly guess the password
\cite{pliam99}, when the ranked list of passwords is known, but the
exact password is not. Guesswork is given by
\[
	G = \sum_{i = 1}^N i P_i,
\]
where $P_i$ is the probability of the password of rank $i$.

Another strategy for guessing passwords, given the distribution,
is to try the common passwords, but to give up when some
fraction $\alpha$ of the distribution has been covered. The mean
number of guesses associated with this is known as the $\alpha$-guesswork,
$G_\alpha$ \cite{pliam99}. Its value is given by
\[
	G_\alpha = \sum_{i = 1}^{r_\alpha} iP_i,
\]
where $r_\alpha$ is the rank of the password when the cumulative
probability of being successful is at least $\alpha$. We will work
with $\alpha = 0.85$, so that we cover most of the distribution,
but avoid the tail.

The Shannon Entropy,
\[
	H = - \sum{P_i \log_{2}P_i},
\]
is a common measure of the number of bits of uncertainty associated
with a random variable. While Shannon Entropy has been used as a
measure of security of password and key distributions, it does not
relate directly to how easy it is to guess a password
\cite{Massey94,malone2005}. R\'{e}nyi entropy, which is a generalization of
Shannon entropy, is given by
\[
	R = \log_2(\sum{\sqrt{P_i}})^2.
\]
It is asymptotically related to the guessability of a password
\cite{ea:aiogaiatsd,malone2004}. The
min-Entropy is also used as a conservative measure of password/key
security \cite{rfc4086}.

\begin{figure}%
{\small
\begin{center}%
\subfloat[hotmail]{\includegraphics[width=\myWidth,height=\myHeight,keepaspectratio]{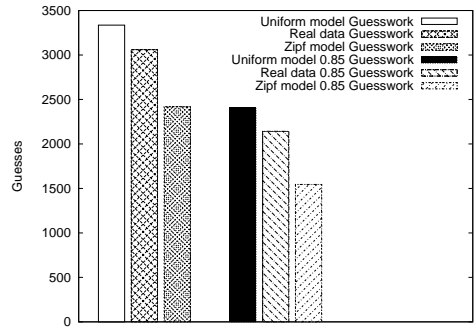}}\\\subfloat[flirtlife]{\includegraphics[width=\myWidth,height=\myHeight,keepaspectratio]{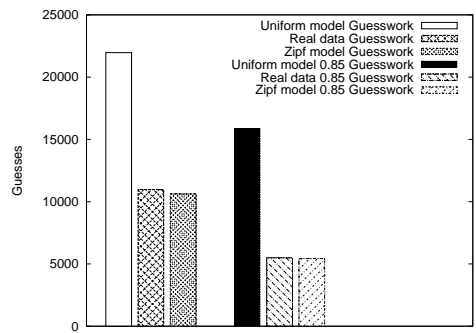}}\\\subfloat[computerbits]{\includegraphics[width=\myWidth,height=\myHeight,keepaspectratio]{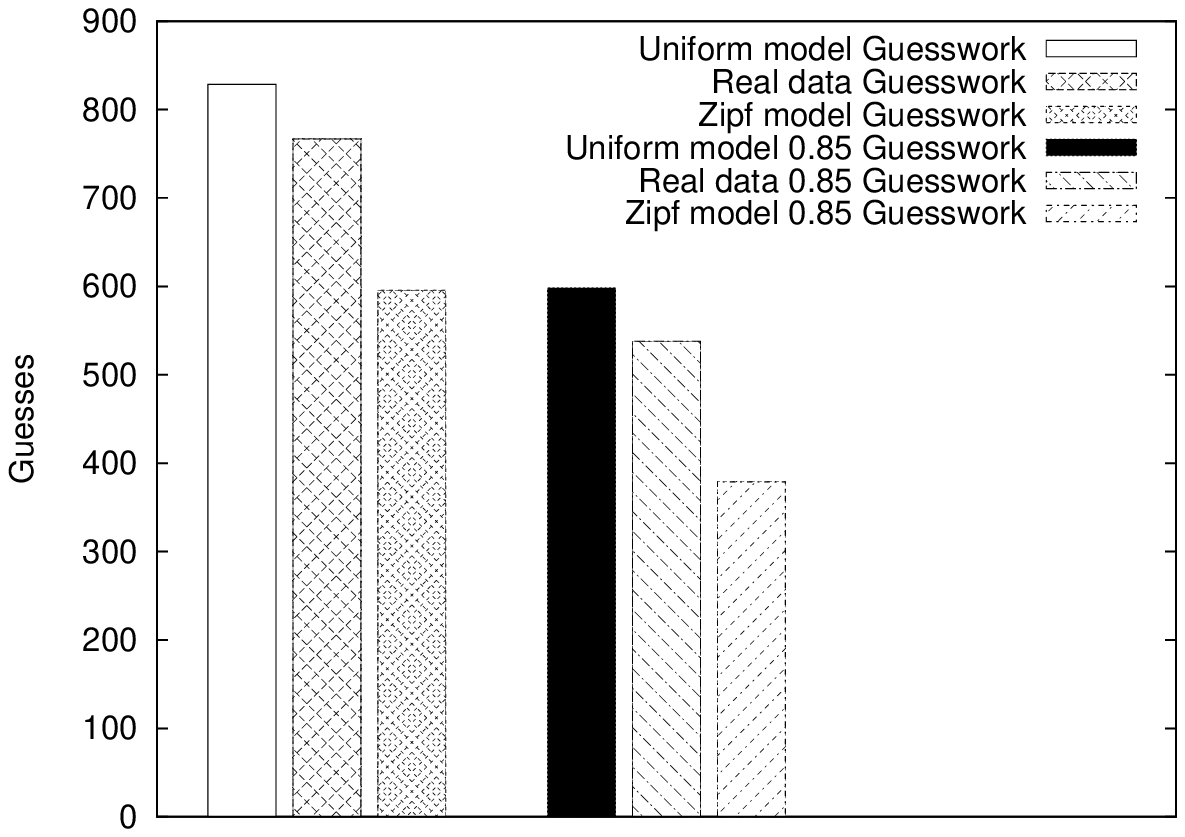}}\\\subfloat[rockyou]{\includegraphics[width=\myWidth,height=\myHeight,keepaspectratio]{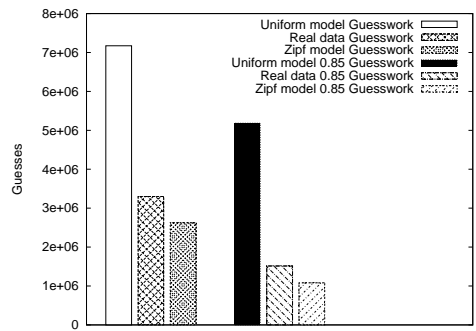}}\\\end{center}%
\caption{Guesswork statistics for Uniform model, Real data and Zipf model.}%
\label{fig:guesswork}%
}
\end{figure}

Figure~\ref{fig:guesswork} compares the guesswork statistics for
the uniform model, the real data and the Zipf model. The three bars
on the left show the guesswork and the three on the right show the
0.85-guesswork. As expected, the guesswork estimates for the uniform
model overestimate the required number of guesses. A relatively
small percentage of the total number of passwords in the hotmail.com
and computerbits.ie lists are shared. This seems to be reflected
in the predictions for guesswork, where the uniform distribution
provides a relatively good prediction for hotmail and computerbits,
while the Zipf model underestimates.  For flirtlife and rockyou,
shared passwords make up a larger percentage of the total passwords,
and the guesswork is far lower than the uniform guesswork, but the
Zipf model provides better predictions, though it still underestimates.

\begin{figure}%
{\small
\begin{center}%
\subfloat[hotmail]{\includegraphics[width=\myWidth,height=\myHeight,keepaspectratio]{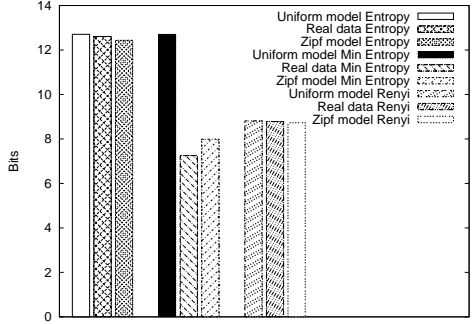}}\\\subfloat[flirtlife]{\includegraphics[width=\myWidth,height=\myHeight,keepaspectratio]{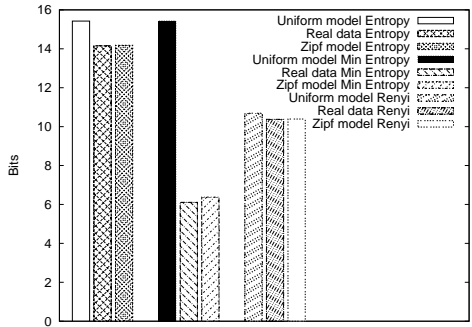}}\\\subfloat[computerbits]{\includegraphics[width=\myWidth,height=\myHeight,keepaspectratio]{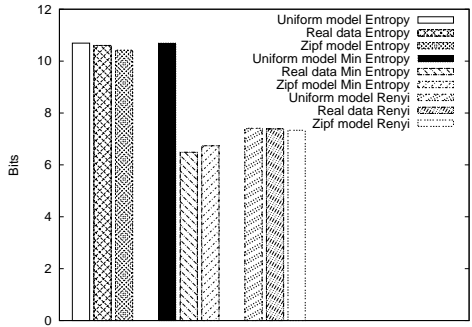}}\\\subfloat[rockyou]{\includegraphics[width=\myWidth,height=\myHeight,keepaspectratio]{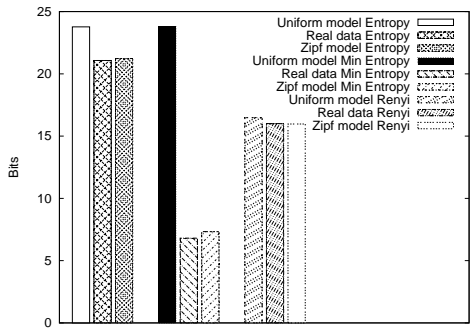}}\\\end{center}%
\caption{Entropy values for Uniform model, Real data and Zipf model.}%
\label{fig:entropy}%
}
\end{figure}

Figure~\ref{fig:entropy} shows the Entropy values for the actual
data and models. Shannon Entropy is shown on the left, min-Entropy
in the middle and R\'{e}nyi Entropy on the right. The Uniform model, again
as expected, tends to overestimate the Entropy. However, for the
R\'{e}nyi Entropy both models and the data seem to give results that
are close together. The Zipf model seems to provide relatively good
approximations in all cases.

\section{The Relation between Distributions}
\label{sec:relativeguess}

We have seen that while the popular passwords in our lists have
things in common (for example, the password 123456), they also show
features specific to the website or service. In Section~\ref{sec:dist}
we also saw that all lists have a number of relatively frequently
used passwords followed by a long tail of uncommon passwords. In
this section, we would like to quantify how much overlap there is
between the passwords in these lists.

\begin{figure}%
{\small
\begin{center}%
\subfloat[hotmail]{\includegraphics[width=\myWidth,height=\myHeight,keepaspectratio]{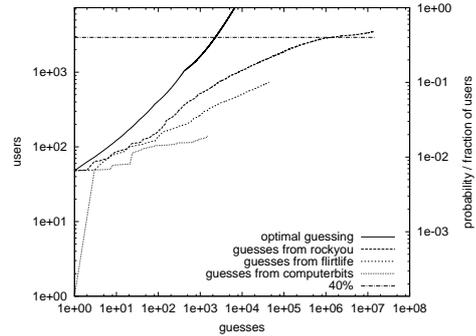}}\\\subfloat[flirtlife]{\includegraphics[width=\myWidth,height=\myHeight,keepaspectratio]{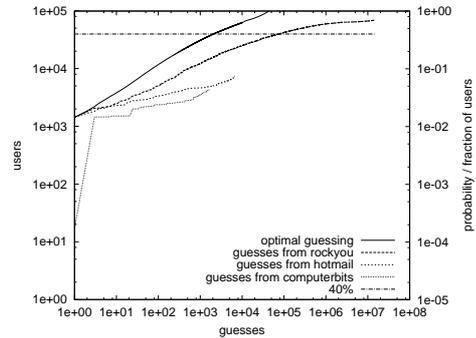}}\\\subfloat[computerbits]{\includegraphics[width=\myWidth,height=\myHeight,keepaspectratio]{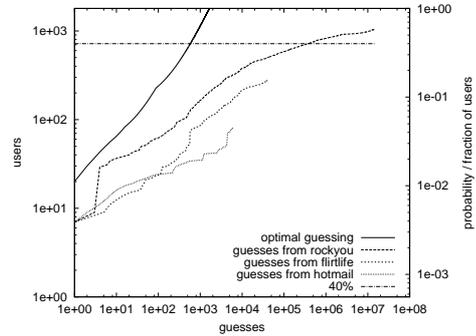}}\\\subfloat[rockyou]{\includegraphics[width=\myWidth,height=\myHeight,keepaspectratio]{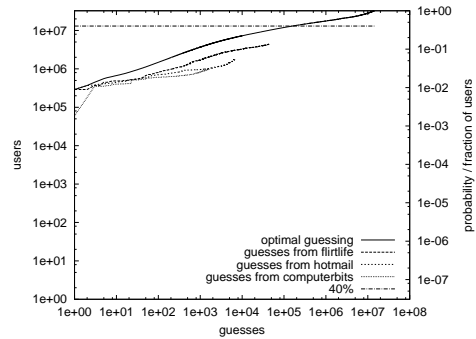}}\\\end{center}%
\caption{Number of users whose passwords are recovered by using the password ordering from one distribution to guess another.}%
\label{fig:relative_guesswork}%
}
\end{figure}

Consider the problem of guessing the password of a randomly selected
user from one of our lists. If we guess the passwords in the order
from most popular to least popular \emph{in that list}, then after
$t$ guesses we will have guessed the passwords used by
\[
	C(t) = \sum_{i = 1}^t f_i,
\]
users. If we guess one password at each trial, guessing in this
order recovers users' passwords as quickly as possible, and is in
this sense optimal. Figure~\ref{fig:relative_guesswork} shows $C(t)$
as a solid line for each of our datasets. The right-hand axis shows
$C(t)/N$, which we can interpret as the probability of successfully
guessing in $t$ guesses or the fraction of users whose passwords
have been guessed. For example, after 100 guesses using the hotmail
data, we have recovered around 400 users' passwords, which represents
a 5\% probability of success against a particular user. Since we
guess in the optimal order, other orderings must recover fewer users and
have a lower probability of success.

If we do not know the optimal order in which to guess the passwords,
we may instead guess them in the optimal order for another reference
data set. Suppose we have a password of rank $i$ in the reference
data set, and it has rank $\sigma(i)$ in the data set being guessed.
If we guess in the order given by the reference data set, after $t$
guesses, we will have guessed the passwords of
\[
	C(t||\sigma) = \sum_{i = 1}^t f_{\sigma(i)},
\]
users, where we assume $f_{\sigma(i)}$ is zero if password $i$ is not
in the list we are guessing. If the ordering of the passwords
by popularity is the same for both lists, then this function
will be $C(t||\sigma) = C(t)$, otherwise $C(t||\sigma) \le C(t)$.

Figure~\ref{fig:relative_guesswork} shows $C(t||\sigma)$ for each of
our lists, when using each of the other lists as a reference.
Consider the situation after 1000 trial guesses. The number of users
whose passwords match one of these 1000 guesses, $C(1000||\sigma)$,
can be seen to vary by almost an order of magnitude, depending on
the list used as a reference. Thus, to guess passwords quickly,
we would like a good reference list.

Observe that for any list, once we have made more than 10--100
guesses, the larger reference lists lead to more successes than
smaller reference lists. This suggests that beyond the most popular
passwords, there may be a more general ordering of passwords that
is more apparent from the larger data sets.  The rockyou list is
quite effective and the top one million rockyou passwords cover
close to 40\% of users in the other lists.

We can apply this directly to a password cracking problem. In
December 2010, the password database from Gawker.com was leaked.
This database did not contain plaintext passwords, but instead
contained hashes of passwords using the well-known DES and Blowfish
password hashing schemes. We can use the words in our list as
guesses in an off-line cracking attack against the Gawker hashes.

\begin{figure}%
{\small
\begin{center}%
\includegraphics[width=\myWidth,height=\myHeight,keepaspectratio]{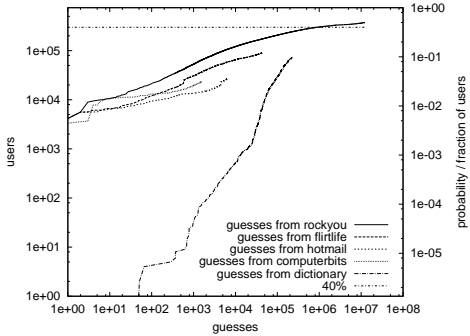}\end{center}%
\caption{Number of users whose passwords are cracked by using the password ordering from different distribution on the Gawker.com password hashes.}%
\label{fig:gawker_relative_guesswork}%
}
\end{figure}

The Gawker dataset contained 748,090 users potentially valid (i.e.
13 character) DES hashes. The hashes use 3844 different salts. A
simple perl script can attempt passwords at a rate of approximately
80,000 trials per hour per core on a modern CPU. As the DES hash
truncates passwords to 8 characters, we truncate long passwords and
reaggregate our previous lists. The number of users whose passwords
were cracked after $t$ trials is shown in
Figure~\ref{fig:gawker_relative_guesswork}.

Again, the large lists provide the fastest recovery of passwords,
and recovers 40\% of users in less than one million trials.  Even
our smaller lists do well, recovering the passwords of more than
10,000 users in around 1,000 trials (less than one minute of CPU
time).

For comparison, in Figure~\ref{fig:gawker_relative_guesswork} we
show the results of using a dictionary in lexical
order as list of guesses. The dictionary is based on the contents
of \texttt{/usr/share/dict/} on Mac OS X, truncated to 8 characters
and sorted using the Unix sort command. This results in a return
on effort that is substantially lower than with ranked password
lists.

Up to now, we have measured the effectiveness of guessing passwords
by counting the number of distinct users whose passwords would have
been correctly guessed after $t$ guesses. An alternative to this
metric, is to look at the number of distinct passwords that have
been correctly guessed after $t$ guesses. In this case, we recover
either zero or one password after each guess.

\begin{figure}%
{\small
\begin{center}%
\subfloat[hotmail]{\includegraphics[width=\myWidth,height=\myHeight,keepaspectratio]{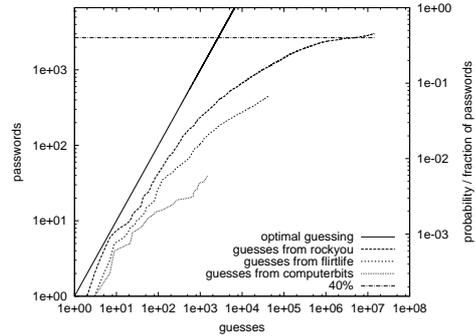}}\\\subfloat[flirtlife]{\includegraphics[width=\myWidth,height=\myHeight,keepaspectratio]{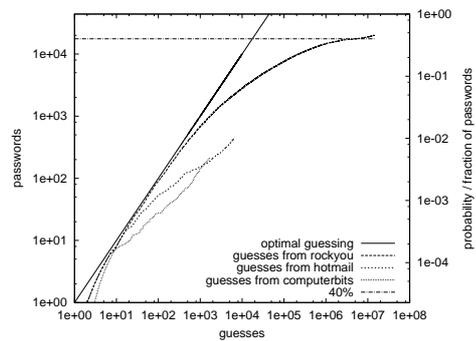}}\\\subfloat[computerbits]{\includegraphics[width=\myWidth,height=\myHeight,keepaspectratio]{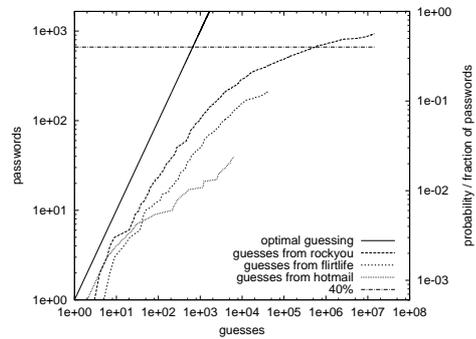}}\\\subfloat[rockyou]{\includegraphics[width=\myWidth,height=\myHeight,keepaspectratio]{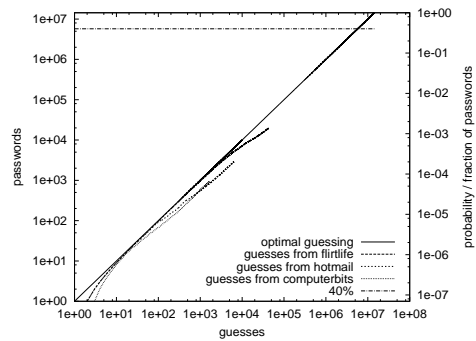}}\\\end{center}%
\caption{Number of passwords recovered by using the password ordering from one distribution to guess another.}%
\label{fig:relative_guesswork_pass}%
}
\end{figure}

Figure~\ref{fig:relative_guesswork_pass} shows the results for our
main lists. The optimal rate at which we can recover passwords is
1 per guess, so we plot the optimal line $y=x$. We see that with
about 500,000--5,000,0000 guesses we can obtain about 40\% of the
passwords, except when guessing rockyou passwords, when the other
lists simply do not have enough guesses to reach 40\%. Despite this,
when guessing passwords from the rockyou dataset, the curves for
the other lists stay close to the optimal line, showing that there
is a good return for each guess.

\begin{figure}%
{\small
\begin{center}%
\includegraphics[width=\myWidth,height=\myHeight,keepaspectratio]{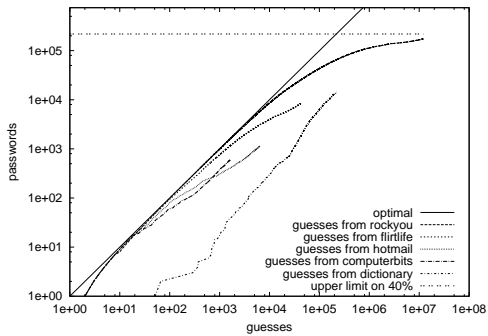}\end{center}%
\caption{Using orderings from each list to crack Gawker password hashes.}%
\label{fig:gawker_relative_guesswork_pass}%
}
\end{figure}

Figure~\ref{fig:gawker_relative_guesswork_pass} shows the results
for guessing the Gawker passwords, in terms of fraction of passwords
recovered. As not all hashes have been cracked and the hashes are
salted, we do not know the total number of distinct passwords.
However, we can upper bound the number by assuming that all uncracked
passwords are unique.

First we note that using other password lists to guess still provides
significantly better return than using a dictionary. Indeed, the
curves stay relatively close to the optimal line for guesses based
on the rockyou data set for between 10 and 10,000 guesses, indicating
a success rate of almost 100\%. After using all 14 million passwords
in this list, we have cracked close to 40\% of the passwords.  The
curve for dictionary words stays a significant distance from the
optimal line, suggesting less than 10\% of dictionary words are
actually used as passwords in the Gawker data set.

In \cite{dellamico2010}, the authors consider various techniques
for generating candidate passwords for guessing/cracking. These
techniques include dictionary attacks, mangled dictionaries and
Markov generators, which can be trained on sample passwords. They
assess these techniques using the fraction of passwords recovered
in three data sets.  They show that in order to recover a substantial
fraction of passwords, say 40\%, the number of required guesses is
over 100 million, unless the password-generating technique is trained
on a similar dataset.

Our results show that the use of a large set of passwords as a
source of guesses seems to offer considerably better returns than
these techniques, being able to recover 40\% of passwords in a less
than 10 million guesses. Of course, once exhausted, a list provides
no more candidate guesses, whereas the mangling and Markov techniques
can theoretically yield unbounded data sets.

\section{Making Password Choices More Uniform}

We have seen that people's choice of passwords is quite non-uniform,
leading to some passwords appearing with a high frequency. The
previous section demonstrated one consequence of this: a relatively
small number of words can have a high probability of matching a
user's password. It has been relatively common practice to ban
dictionary words or common passwords (e.g. the Twitter banned
password list \cite{twitter}), in an effort to drive users away
from more common passwords.

In fact, if password choices were uniform (over a large set of
passwords) some attacks based on the existence of common passwords
become ineffective. Based on this, a scheme was suggested in
\cite{Schechter10} which prevents uses from choosing particular
passwords when they become relatively too popular, in an effort to reduce the
non-uniformity of the password distribution.

However, there are well-known schemes that use the output from a
random generator
and manipulate it to achieve a different distribution. For
example, the Metropolis-Hastings scheme allows us to generate a
desired distribution $P(.)$ by probabilistically accepting/rejecting
the steps of a Markov chain $Q(.,.)$. It has a useful feature that
the density of the desired distribution does not need to be known,
as long as a function proportional to the density is known.

The basic Metropolis-Hastings scheme is as follows:
\begin{enumerate}
\item Set $t \leftarrow 0$, choose $x^t$.
\item Generate $x'$ with distribution $Q(x', x^t)$.
\item With probability
	\[ \min\left(1,\frac{P(x')}{P(x)} \frac{Q(x^t, x')}{Q(x',x^t)}\right) \]
	go to step 4 (accept), otherwise return to step 2 (reject).
\item $t \leftarrow t + 1$, $x^t \leftarrow x'$.
\end{enumerate}
where the terms of the sequence $x^t$ are the output. Usually, the
initial outputs are discarded, to wash out the initial choice of
$x^0$ and allow the sequence to move closer to its stationary
behaviour. This is sometimes referred to as \emph{burn in}.

We can apply such a scheme to produce a more uniform choice of
passwords. Our desired distribution $P(.)$ will be uniform over all
passwords that users are willing to use, so $P(x')/P(x) = 1$. Next,
we suppose that if
a user is asked to choose a password that they choose it independently
of previous choices with probability $Q(.)$. Since we do not know
$Q(.)$, we will estimate it in an online fashion, by tracking the
frequencies $F(.)$ with which users suggest passwords.

This suggests the following scheme when users select a password.
\begin{enumerate}
\item Choose a password $x$ uniformly from all previously seen passwords.
\item Ask user for a new password $x'$.
\item Generate a uniform real number $u$ in the range $[0,F(x')]$
	and then increment $F(x')$. If $u \le F(x)$ go to step 4
	(accept), otherwise return to step 2 (reject).
\item Accept use of $x'$ as password.
\end{enumerate}
This scheme aims to generate a uniform distribution via users'
choices using the Metropolis-Hastings scheme. There are a few things
to note. First, by choosing $x$ uniformly over all seen passwords,
we aim to avoid the need for a burn-in period, because we begin
with a choice drawn from the distribution we want. Second, the password
$x$ is never actually used, only its frequency $F(x)$, so we can
use $0$ for $F(x)$ if the scheme has seen no previous passwords.
Finally, the scheme learns $F(.)$ on-line, so the more password
choices it sees, the better we expect it will be at making choices
uniform.

\begin{figure}%
{\small
\begin{center}%
\includegraphics[width=\myWidth,height=\myHeight,keepaspectratio]{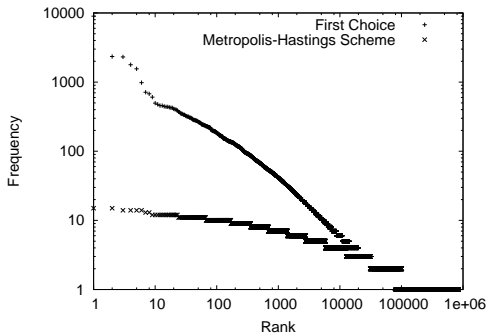}\end{center}%
\caption{Plot of rank vs. frequency for users generated from rockyou dataset, with free choices and with the Metropolis-Hastings scheme.}
\label{fig:levellers}%
}
\end{figure}

We implemented this scheme and tested it by choosing passwords from
the rockyou dataset, where the probability a password was selected
was proportional to its frequency in the dataset. We generated
passwords for 1,000,000 users, with the Metropolis-Hastings scheme
and, for comparison, when users were free to choose any password.
The results are shown in Figure~\ref{fig:levellers}.

We see that with the Metropolis-Hastings scheme, the distribution
is much more uniform, and the frequencies of the most common passwords
have been reduced from over 1,000 to just over 10, a reduction of
more than two orders of magnitude. One concern with this scheme is that is
may reject a user's choice of password many times, and frustrate
the user. However, over these trials, users are asked on average
for 1.28 passwords with a variance of 0.61.

We note that this scheme has some similarities to the scheme in
\cite{Schechter10}. Both schemes must store frequency information
about passwords. To avoid storing the passwords in plain text, one
could store the frequency of hashed passwords. Our scheme, however,
stores information about the frequency with which users choose a
password rather than the frequency information for passwords in
use. This has some advantages if the frequency table is stolen by
an attacker, as even if the hashes can be cracked, frequent choices
are not so commonly used because of the Metropolis-Hastings scheme.

Rather than using a simple frequency table,
both these schemes can be implemented with count-min sketches.
This is an efficient data structure that stores estimates of frequencies.
As described in detail in \cite{Schechter10}, there are advantages
to storing the information in a sketch, particularly if the sketch
is stolen by an attacker. This is because it uses multiple hash
functions with a smaller output space, leading to false-positives
if an attacker tries to identify high-frequency passwords.

One difference in these schemes is that the Metropolis-Hastings
algorithm aims to make the whole distribution more uniform, rather than
limit the frequency of the most popular passwords. As we saw in
Section~\ref{sec:relativeguess} mid-ranked passwords, say from rank
10--1,000, are important for increasing the success rate when guessing a
user's password; these guesses increase the success probability
from a fraction of a percent up to a few percent.
By moderating the frequency of these passwords, we reduce the
effectiveness of attacks that use both high- and mid-ranked passwords.

We also note that while this scheme learns the password frequency
distribution on-line, it could also be initialised using a known
list of password frequencies. While we chose to target a uniform
distribution, this scheme could also be combined with a list of
banned passwords (by setting the desired frequency $P(x)$ to be
zero) or implement a soft-ban on some passwords (by reducing $P(x)$
for those passwords).

\section{Discussion}

We have seen that while Zipf's law is not an exact match, it seems
to provide a passable description of the frequencies with which
passwords are chosen. Estimates of the parameter $s$ are considerably
less than one.  While this might be interpreted as indicating a
strongly heavy tailed behaviour, another interpretation is that as
$s \rightarrow 0$ the distribution becomes uniform, which is actually
desirable for passwords. These observations may be of use to algorithm
designers, for dimensioning data structures or even taking advantage
of the relatively heavy-tailed nature of users' choices.

We also see that fitting a distribution provides relatively good
approximations of the Shannon Entropy, guesswork and other statistics
that are of interest when assessing a password distribution. Using
a uniform model, where all passwords are equally likely, provides
reasonable approximations for the data sets with smaller $s$, but
provides a poor estimate of min-Entropy.

We have seen that demography of the userbase choosing the passwords
can be evident in the most popular passwords, and even the name of
the website is a likely password. Some sites, for example Twitter,
have noticed this and
implement banned password lists \cite{twitter}, which includes many
of the more common passwords, including the name of the site. This
also gives weigth to the advice that administrators checking the
security of passwords in use at their site using password cracking
software should include custom dictionaries including locally used
terms.

The Zipf distribution decays relatively slowly, so we expect there
to be a large number of relatively commonly chosen passwords. We
investigated if these passwords change much from one list to another.
We see that this is the case, and that while not optimal, larger
lists provide good guidance about the ranking of passwords in other
lists. We've demonstrated that this can provide a significant speedup
in guessing or cracking passwords with moderate numbers of guesses,
particularly over simple dictionary attack, but also over a range of
the guess-generating techniques described in \cite{dellamico2010}.

An attacker who has collected leaked passwords from a collection
of websites has a useful starting point for cracking a password. If
a hashed password is exposed, the time for an attacker to try, say,
20 million passwords is relatively small, even on a single CPU.  We
note that this adds some extra weight to the advice that reusing
passwords between websites is a risk, even if there is no way for
an attacker to identify which pairs of users are common to the
websites. This is because if just one site stores the password in
an unhashed format and that password is leaked, then it facilitates
the subsequent cracking of that password on a system where the
passwords are hashed.

Banning more commonly chosen passwords may result in a more even
spread of password in use. Interestingly, we saw that most English
dictionary words are not necessarily common passwords: out of more
than 220,000 dictionary words, less than 15,000 appeared as passwords
in the Gawker data set. We proposed a scheme based on the
Metropolis-Hastings algorithm that aims to generate more uniform
password choices, without having to know a list of common passwords
in advance. A basic implementation of this is relatively straight
forward, and could be easily incorporated into a PAM module
\cite{Samar96}.

\section{Conclusion}

We have seen that a Zipf distribution is a relatively good match
for the frequencies with which users choose passwords. We have also
seen that the passwords found in the lists that we have studied
have relatively similar orderings. Consequently, passwords
from one list provide good candidates when guessing or cracking
passwords from another list. Finally, we present a scheme that can
guide users to distribute their passwords more uniformly.

\emph{Acknowledgment:} The authors would like to thank Ken Duffy
for thought-provoking comments and Dermot Frost for the offer of
spare CPU cycles.

\bibliographystyle{plain}
\bibliography{Report}

\end{document}